\documentstyle[aps,epsfig]{revtex}
\begin {document}
\title {Crystallization of a supercooled liquid and of a glass -- Ising model
approach} 
\author{A.~Lipowski$^{1)}$ and D.~Johnston$^{2)}$}
\address{
$^{1)}$ Department of Physics, A.~Mickiewicz University,
61-614 Pozna\'{n}, Poland\\
$^{2)}$ Department of Mathematics, Heriot-Watt University
EH14 4AS Edinburgh, United Kingdom\\}
\maketitle
\begin {abstract}
Using Monte Carlo simulations we study crystallization in the
three-dimensional Ising model with four-spin interaction.
We monitor the morphology of crystals which grow after
placing crystallization seeds in a supercooled liquid.
Defects in such crystals constitute an intricate and very stable network which
separate various domains by tensionless domain walls.
We also show that the crystallization which
occurs during the continuous heating of the glassy phase takes place at a
heating-rate dependent temperature.
\end{abstract}
\section{introduction}
Although of great experimental and technological importance and the subject
of intensive experimental, theoretical and numerical research, crystallization
is still far from completely understood~\cite{CRYSTAL}.
One of the main difficulties which hinders comparison of numerical results with
experiments and phenomenological theories is the fact that most
realistic, off-lattice  models, which can be studied using molecular dynamics
simulations, still constitute an enormous computational challenge. 
Typically, computationally accessible systems contain only up to $10^4$ atoms
and can be studied during a time interval which is several orders of
magnitude shorter than would be required for comparison with experiment. 
Recently, however, important developments in this field have taken place and
some aspects of the
numerical simulations have been successfully compared with
experiment and theory~\cite{HUITEMA}.

A possible alternative to study crystallization might be to examine lattice
models.
Although less realistic, such models are usually much easier to simulate and
in some cases analytical approaches can also be used.
Since crystallization typically occurs during a slow cooling of supercooled
liquid, an appropriate model, depending on the thermodynamic parameters and
its history should exhibit crystal, liquid or glassy characteristics. When
the cooling is too fast liquid does not crystallize but collapses into the
glassy phase.

At first sight it seems that a natural candidate for such a model might be
a standard nearest neighbour interaction Ising model.
Crystal and liquid phases can be then easily related with low- and
high-temperature phases of this model.
However, the relatively fast dynamics of standard Ising models 
precludes the
existence of a glassy phase.
In other words, in such models the liquid usually quickly evolves toward the
crystal phase when the temperature is below the transition point and does 
not get
trapped in the glassy phase.

Such behaviour is typically observed in studies of most Ising-like models
but fortunately there are also some exceptions
for which the dynamics {\it can} be
very slow, similar to some glassy systems.
Models of this kind might be infinite-dimensional (mean-field),
and thus exactly solvable with respect to some quantities~\cite{BERNASCONI},
or finite-dimensional with certain properties typical of glassy
dynamics~\cite{SHORE}.

A promising, finite-dimensional model of glassy systems is a three-dimensional Ising
model with the four-spin, plaquette, interaction.
Recent work has shown that this relatively simple system exhibits a number of properties
characteristic of glasses including strong metastability~\cite{LIPDES1} and slow
ordering under cooling~\cite{LIPDES2}.
In addition,
certain time dependent correlation functions also behave in a way 
typical of glassy
systems~\cite{SWIFT}.
Since this model is homogeneous (i.e., does not contain quenched disorder) it
has a crystal phase and under certain conditions it should crystalize.
An extremely strong metastability of supercooled liquid in this
model suppresses, however, (within computationally accessible system size and
simulation time) spontaneous nucleation and crystallization was not observed.
However, keeping a certain fraction of spins fixed, one enhances
an ordered structure and under slow cooling the system evolves
toward the crystal phase~\cite{LIPDES2}.

Since the four-spin model has the properties which are required to model
crystallization it would be desirable to examine this process under conditions
which more closely resemble experimental realizations.
Such an examination is an objective of the present paper.
In section II we introduce the model and briefly describe its properties.
In section III we examine constant-temperature crystallization.
To enhance crystallization we use some
crystallization seeds rather than fixing a certain fraction of spins
dispersed throughout the system.
We then monitor configurations of the system to examine the
morphology of
the growing crystals.
We observe that for small supercooling the growth of crystal is surface-tension
dominated while for larger supercooling irregular crystals are formed.
We also examine the distribution of defects and their stability.

In section IV we examine crystallization of glass under continuous heating.
Without crystallization seeds our model never crystallizes and enters the liquid phase 
at the equilibrium transition temperature.
However, when the crystallization is enhanced we observe crystallization which
takes place at the heating-rate-dependent temperature.
We also estimate the critical heating rate and show that this is probably
larger
than the corresponding critical cooling rate.
A similar assymetry between cooling and heating processes was recently observed
experimentally~\cite{SCHROERS}.
Section V contains our conclusions.
\section{Model and its basic properties}
Our model is the three-dimensional (cubic) Ising model with the four-spin
(plaquette) interaction,
which is described by the Hamiltonian
\begin{equation}
H=-J\sum_{[i,j,k,l]} S_iS_jS_kS_l,
\label{e1}
\end{equation}
where summation is over all elementary plaquettes and $S_i=\pm 1$.
In the following we put $J=1$.
Recently, model (\ref{e1}) was studied in the context of modelling of
conventional glasses~\cite{LIPDES1,LIPDES2,SWIFT,LIPDES3}. 
Morover, this model and its extensions were used in certain lattice gauge
theories~\cite{SAVVIDY}.

Model (\ref{e1}) has strongly degenerate ground state.
Of course a ferromagnetic configuration minimizes the Hamiltonian,
but one can easily notice that any configuration obtained from a ferromagnetic
one by flipping an entire plane of spins also belongs to the ground state.
Elementary counting shows that degeneracy of the ground state equals $8^L$,
where $L$ is the linear system size.
Let us notice that lamellar structures, for example made of
parallel ferromagnetic layers also belongs to the ground state and
in the next
section we discuss the role of such structures in formation of defects.

Results of various Monte Carlo simulations for this model can be
summarized as follows:
The model has two equilibrium phases: high- and low-temperature phases
referred to as liquid and crystal.
First order transition between them takes place at temperature
$T=T_c\sim 3.6$.
However, this transition is screened by very strong metastability of the model
both upon cooling and heating.
Of particular interest to our study is metastability of liquid: in the
temperature range $3.4<T<3.6$ crystal is thermodynamically stable since
its free energy is lower than that of liquid, nevertheless even very long 
simulations are not sufficient to transform liquid into crystal.
Only below $T=T_g\sim 3.4$ does the liquid collapse into the glassy phase.
Various characteristics of glassy dynamics have been shown
to be present such as cooling-rate
effects~\cite{LIPDES2} and time-dependent correlation functions~\cite{SWIFT}.

An important property of glasses is their slow dynamics.
Monte Carlo simulations show that model (\ref{e1}) also has slow low
temperature dynamics~\cite{LIPDES1}.
In particular for $T<T_g$ the excess energy (above the equilibrium value) $\delta E$ of
the random quench decays most likely logarithmically in time,
which should be contrasted with the typical nonconservative dynamics
decay $\delta E \sim t^{-1/2}$~\cite{BRAY} ($t$-time).
The slow dynamics of model (\ref{e1}) was  
conjectured to be due to diverging energy
barriers which are generated during the evolution of the quench~\cite{LIPDES1}.

It was observed that even upon very slow continuous cooling the liquid
always collapses into the glass (and not crystal)~\cite{LIPDES2}.
To transform liquid into crystal one has to enhance crystallization.
One possibility, which was already used~\cite{LIPDES2} is to fix certain
fraction of spins as e.g., 'up'.
The fixed spins were randomly distributed throughout the system.
Such a procedure strongly favours the ferromagnetic ground state and it was
observed that upon slow cooling the liquid indeed crystallizes.
However, such a procedure differs considerably from typical experimental setups
used in the crystal growth.
In particular the resulting crystal phase is basically homogeneous and does
not contain any large scale defects.
The objective of the next two sections is to examine the model under
conditions which are more representative of real crystal growth experiments.
\section{Constant-temperature growth}
First we examine the constant-temperature crystallization of supercooled
liquid (i.e., $T_g<T<T_c$). 
As an initial configuration we take a random configurations of spins
and to enhance crystallization we add a
centrally placed crystal seed of a cubic cluster of 'up' spins.
This configuration then evolves according to the standard Metropolis
algorithm~\cite{BINDER}.
When the size of the seed is sufficiently large the surrounding liquid will
gradually crystallize.

Due to the strong degeneracy of the ground state many varieties of domains are
formed during the evolution of the model and visualization of the process
constitutes a nontrivial problem. 
To overcome  this difficulty we decided to monitor plaquettes which are
'unsatisfied', i.e., for which $S_iS_jS_kS_l=-1$.
In such a way we monitor plaquettes contributing to the excess energy $\delta
E=E-E_o$, where $E_o$ is the energy of the ground state.

\begin{figure}
\begin{center}
\epsfig{file=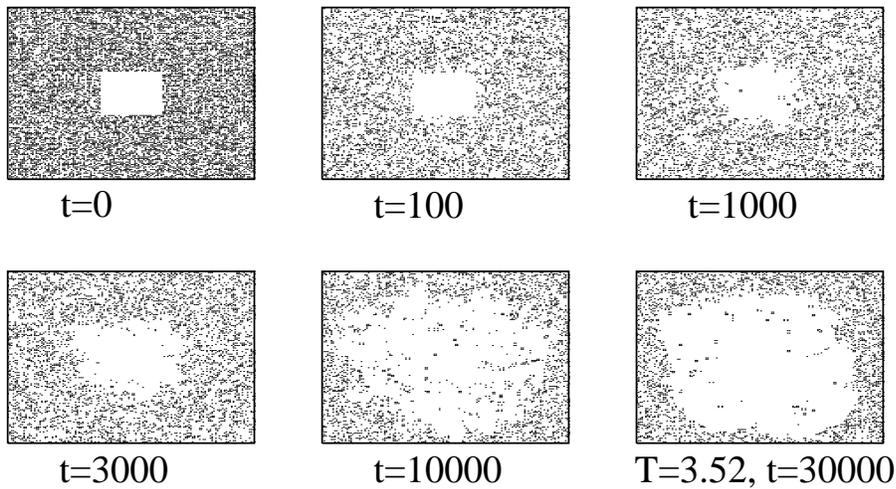,width=15cm}
\end{center}
\caption{
A planar section showing the distribution of 'unsatisfied' plaquettes.
Simulations were done at temperature $T=3.4$ and for the system size $L=140$ with
the crystal seed of the size $L'=35$.
Only for bottom right plot the temperature was $T=3.52$.
Periodic boundary conditions in all directions are used.
}
\label{f1} 
\end{figure}

A typical evolution of our system is shown in Fig.~\ref{f1}.
For $t=0$ unsatisfied plaquettes are found only in the surrounding liquid.
After a short time ($t=100$) the central seed is basically unchanged but the
density of unsatisfied plaquettes is diminished.
This is because, due to the relatively fast
dynamics of liquid, the random configuration of 
the external spins has relaxed to a typical
liquid configuration, which apparently contains smaller concentration of
unsatisfied plaquettes than a genuinely random configuration.
Further growth leads to a relatively irregular crystal, which is
related with the large supercooling, $T=3.4$.
For smaller supercooling ($T=3.52$) the resulting crystal (also shown in Fig.~\ref{f1})
is more regular and with sharper boundaries.

\begin{figure}
\begin{center}
\epsfig{file=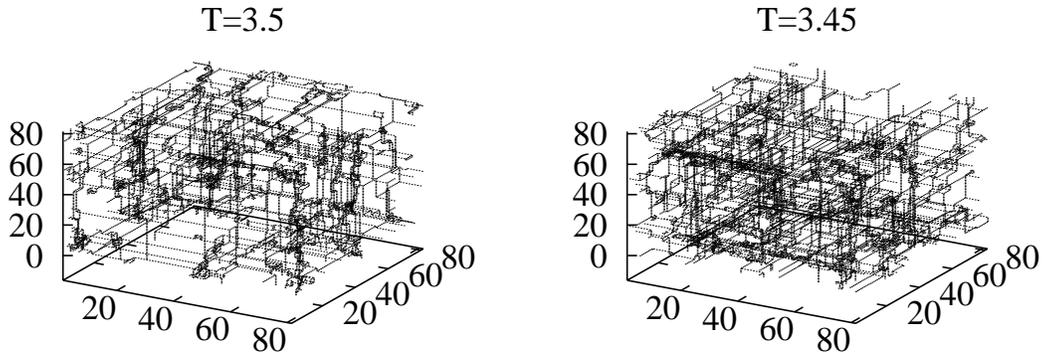,width=15cm}
\end{center}
\caption{
A distribution of 'unsatisfied' plaquettes.
Simulations were done for a constant temperature and for the system size
$L=100$ with the crystal seed of the size $L'=25$.
(Only a portion of the system is shown.)
Simulation time was larger than the time needed for the crystal to fill the whole
lattice.}
\label{f2} 
\end{figure}

In Fig.~\ref{f1} one can see that the crystal contains a certain fraction of
unsatisfied plaquettes.
An important question is
what is the structure of these excitations: are they point-like excitations,
which are basically thermal fluctuations, or are they large
size excitations caused by the complicated dynamics of the model?
To answer this we looked at the three-dimensional structure of
growing crystal (let us note that Fig.~\ref{f1} shows only two-dimensional
cross-sections).
Our results, shown in Fig.~\ref{f2}, strongly suggest that the latter
possibility holds.
Indeed, especially for $T=3.5$ one can see that the unsatisfied plaquettes are 
not randomly scattered but constitute an intricate network.

At this point we recall that in model~(\ref{e1}) tensionless structures can be
formed~\cite{LIPDES2,LIPDES3}.
For example, when a cubic cluster of linear size $L'$ of 'down' spins is surrounded by
'up' spins then the excess energy of such a configuration scales linearly with $L'$ and
not as $L'^2$ as for ordinary two-spin Ising model.
Such scaling is due to the 
fact that unsatisfied plaquettes in this case are only
those that are located at the edges of the cubic cluster 
and not on its surface as in the standard Ising model.
The linear nature of the
excitations in Fig.~\ref{f2} confirms that the resulting crystal is
composed of various domains separated by tensionless domain walls.

It is well known that metastability, 
which is an important property of our model, 
appears also in other models.
For example, let us consider the ferromagnetic two-spin Ising
model below its critical temperature in a phase with positive magnetization
Applying a negative magnetic field, the majority phase becomes
metastable~\cite{RIKVOLD}.
When the magnetic field is weak this metastability might be quite strong and
by placing a crystallization seed (i.e., the negatively magnetized cluster of
spins) we can observe a gradual but slow growth of the thermodynamically  stable
phase.
Using a similar approach a number of interesting results
concerning crystal growth have already
been obtained~\cite{SHNEIDMAN} with the ferromagnetic two-spin Ising
model. 
Why then, it
is natural to ask,
do we study a similar phenomenon in a more complicated four-spin 
Ising model? 
In our opinion,
there are certain reasons to believe that crystallization as observed in a
four-spin model~(\ref{e1}) is more realistic. 
First, similarly to real
crystals, our crystals are imperfect. 
These defects appear because
crystallization, which takes place on different faces  leads to formation of
different domains. 
This effect is caused by the strong degeneracy of the
ground state in the model~(\ref{e1}).
In particular, since lamellar structures are also ground state configurations,
we find that on a one side of the ($+$) seed we might have accumulation of 
($+$)
spins while the other side of the seed might accumulate ($-$) spins.
Such growth inevitably produces defects which must appear on the junctions
of these different domains.
This effect is clearly absent in the two-spin Ising model: due to a double
degenerate ground state a growing crystal is basically homogeneous.

\begin{figure}
\begin{center}
\epsfig{file=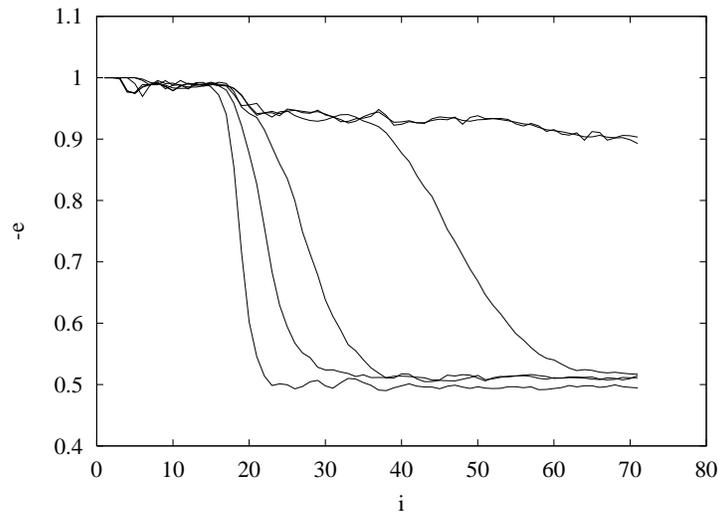,width=10cm,angle=-0}
\end{center}
\caption{
The inverse energy/plaquette $-e$ as a function of a distance of the shell from the
centre $i$. 
Initial configuration as in Fig.~\ref{f1}.
Simulations were made for $T=3.4$.
As the time proceeds one can see a crystallization front moving from left ($t=100$) to
the right ($t=50000$).
Let us notice that two late-stage curves for $t=20000$ and $t=50000$ are almost
indistinguishable which indicates that the resulting pattern of defects almost does not
change in time.}
\label{shells} 
\end{figure}

It is also important to note that this network of defects is very
stable.
We monitored the distribution of energy in cubic shells centered at the centre
of the crystal seed
and the results, averaged over the number of spins in a given shell, for several
times $t$ are shown in~Fig.~\ref{shells}.
One can see the  
propagation of the crystallization front from the centre outward.
Once it reaches size of the system ($t\sim 20000$) the profile of $-e$ stabilizes at
values smaller than unity which confirms formation of a stable network of unsatisfied
plaquettes.
Let us notice that the late stage profile of $-e$ is also a decreasing function of the
distance $i$ from the origin
which indicates that density of defects in 
the resulting crystal increases with $i$. 

The stability of this network of defects is most likely due to energy barriers
which appear in this model.
We have previously conjectured that these
are also responsible for the slow
kinetics of the model below the glassy transition~\cite{LIPDES1,LIPDES2}.
\section{heating of glass}  
Depending on the speed of the process, 
the cooling of  liquid is an important technique used
to produce crystals or glasses.
A parallel technique is based on the heating of glass.
Recently, this technique has been
applied to certain metallic glasses, which
are rather bad glass formers~\cite{SCHROERS}. 
Similarly to the cooling process, there is a critical heating rate which separates the
slow and the fast heating regimes.
Only slow heating of glass leads to crystallization while fast heating 
transforms glass directly into liquid.

We performed a series of simulations to check whether model~(\ref{e1}) exhibits similar
behaviour.
To prepare an initial glassy configuration we quenched a random configuration of spins
at $T=0$ and let the system relax until the system reached the local minimum-energy
configuration.
Such a configuration was subsequently heated with the temperature changing linearly in
time:
\begin{equation}
T(t)=rt,
\label{e2}
\end{equation}
where $r$ is the heating rate.
(The unit of time is defined as a time needed for a single, on average, update of each
spin).

The results of our simulations are shown in Fig.~\ref{f4}.
One can see that within computationally accessible heating rates 
the crystallization of
the glass was not observed.
This data suggest that for infinitely slow heating the temperature of 
melting of the glass
approaches the equilibrium transition temperature at $T=3.6$.

To enhance crystallization we used the simplest approach
of fixing a certain fraction of
spins as 'up'.
As shown in Fig.~\ref{f4} this dramatically changes the behaviour and for slow heating the 
glass crystallizes.
The observed temperature of crystallization depends sensitively on the
heating rate $r$.

In our simulations we fixed 5\% of spins which is the same amount as during the
cooling~\cite{LIPDES2}.
Earlier simulations indicated that for such a fraction of fixed spins the critical
cooling rate $r_c$ satisfies the inequality: $0.0002<r_c<0.0005$.
Results in Fig.~\ref{f4} suggest that for the critical heating rate $r_h$ we have the
bound $0.0005<r_h<0.001$.
These estimates suggest that $r_c<r_h$.
Let us notice that a similar assymetry of this two processes is also observed
experimentally~\cite{SCHROERS}.

\section{conclusions}
In the present paper we have
shown that the four-spin Ising model can be used to model
crystallization.
We observed crystal growth from the supercooled liquid at constant temperature
and found that
crystals of different morphology were obtained depending on how supercooled the
liquid was.
We also examined the structure of defects in the resulting crystals and showed that they
form a very stable network which provides a tensionless separation of different domains
of the crystal.
We also examined the evolution of the glassy phase our model under continuous
heating.

Crystallization  is of course a very complex phenomenon which involves diffusion,
adsorption at the growth surface, crystal
growth and sometimes also additional processes~\cite{VERE}. 
Each of these processes is  complicated and to develop some 
understanding 
one has to introduce some simplifying assumptions. 
As a result certain aspects
of crystallization can be studied using phenomenological models like the
Swift-Hohneberg model~\cite{SWIFTHOH} or Kolmogorov-Avrami  model~\cite{KOLM}.
In principle, however, all aspects of the crystallization process are
determined by the underlying microscopic dynamics of the model. 
Evolution of our model is driven by the Metropolis dynamics which is a 
standard
dynamics for Ising-type models.
In this respect our method is similar to the molecular
dynamics simulations and
although our model is less realistic than off-lattice models, it is computationally much
less demanding.
We consider the results presented here
as rather preliminary 
but they clearly indicate that more detailed
studies 
of the four-spin Ising model
are warranted which can hopefully clarify other aspects of crystallization.
\acknowledgements { This work was partially supported by the KBN grant 5 P03B
032 20, the EC IHP network  ``Discrete Random Geometries: From Solid
State Physics to Quantum Gravity'' {\it HPRN-CT-1999-000161} and the
ESFnetwork ``Geometry and Disorder: From Membranes to Quantum Gravity''.} 

\begin{figure}
\begin{center}
\epsfig{file=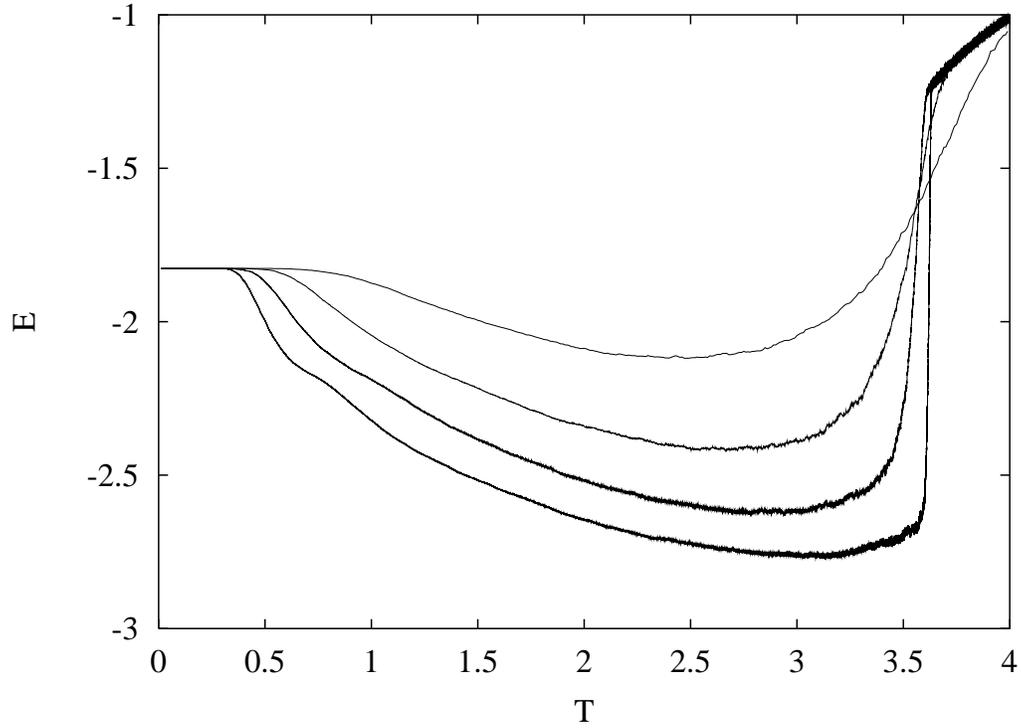,width=10cm,angle=-90}
\end{center}
\caption{
The energy of the model as a function of temperature during a continuous heating of the model
($L=60$) with the heating rate (from top to bottom): 0.01, 0.001, 0.0001, and 0.00001.
Initial configuration is obtained by quenching a random sample at $T=0$ and relaxing the
system until the system reaches a stationary state (i.e., a local energy minimum).
}
\label{f3} 
\end{figure}
\begin{figure}
\begin{center}
\epsfig{file=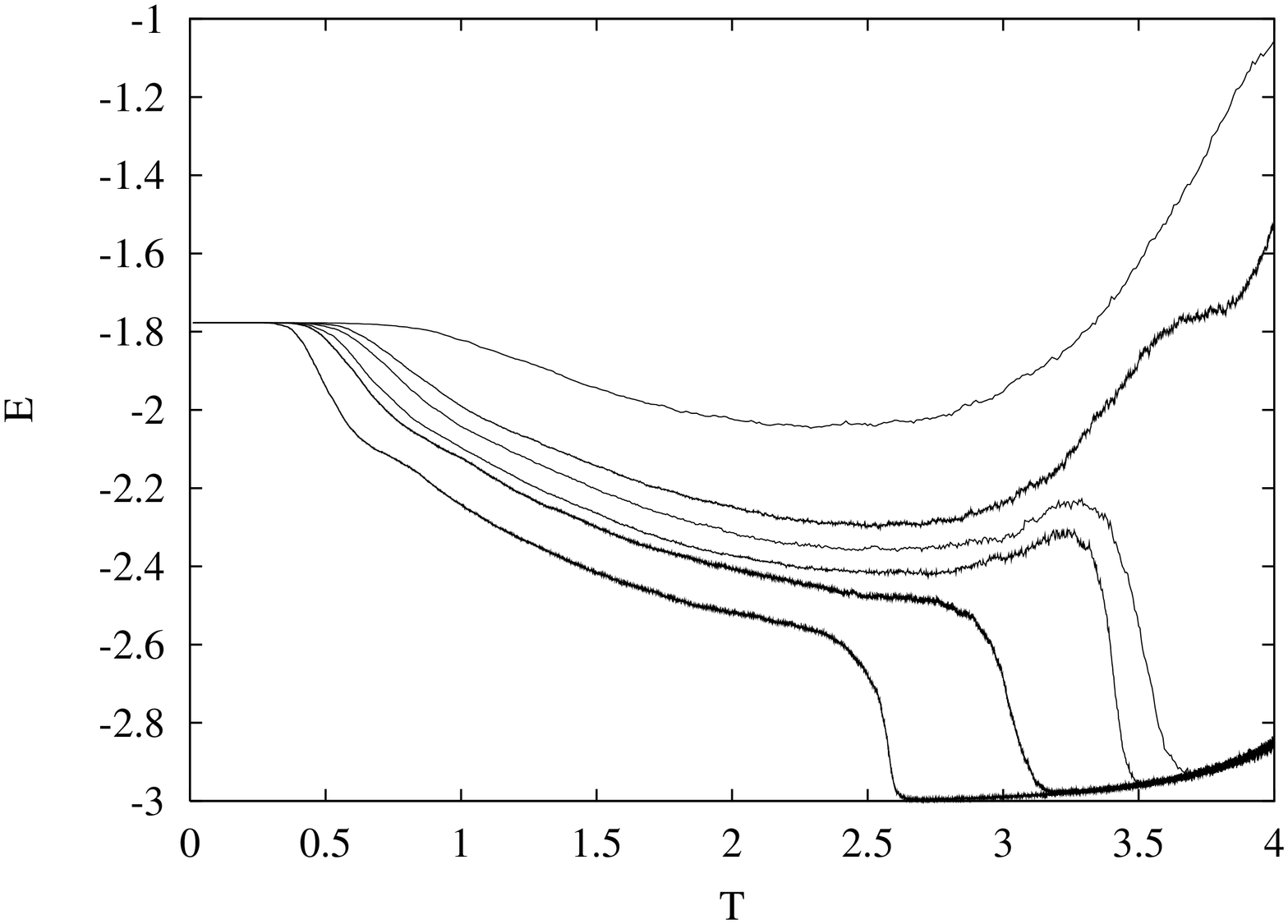,width=10cm,angle=-90}
\end{center}
\caption{
The energy of the model as a function of temperature during a continuous
heating of the model ($L=60$) with $5\%$ of spins being fixed (all-up) and
with the heating rate (from top to bottom): 0.01, 0.001, 0.0005, 0.0002,
0.0001, and 0.00001. 
Initial configuration is the same as in Fig.~\ref{f3}.
For $r\leq0.0005$ the system crystallizes and its energy in the
high-temperature part is the same as during heating of one of the
ground-state configurations.} 
\label{f4} 
\end{figure}
\end {document}